# Imbalance Knowledge-Driven Multi-modal Network for Land-Cover Semantic Segmentation Using Images and LiDAR Point Clouds

**Yameng Wang [a], Yi Wan [a, *], Yongjun Zhang [a, *], Bin Zhang [a], and Zhi Gao [a]**

a    School of Remote Sensing and Information Engineering, Wuhan University, Wuhan 430079, China
*    Corresponding authors.
E-Mail: ymw@whu.edu.cn (Y. Wang), yi.wan@whu.edu.cn (Y. Wan), zhangyj@whu.edu.cn (Y. Zhang), bin.zhang@whu.edu.cn (B. Zhang), gaozhinus@whu.edu.cn (Z. Gao)

**Abstract:** Despite the good results that have been achieved in unimodal segmentation, the inherent limitations of individual data increase the difficulty of achieving breakthroughs in performance. For that reason, multi-modal learning is increasingly being explored within the field of remote sensing. The present multi-modal methods usually map high-dimensional features to low-dimensional spaces as a preprocess before feature extraction to address the nonnegligible domain gap, which inevitably leads to information loss. To address this issue, in this paper we present our novel **I**mbalance **K**nowledge-**D**riven Multi-modal **Net**work (**IKD-Net**) to extract features from raw multi-modal heterogeneous data directly. IKD-Net is capable of mining imbalance information across modalities while utilizing a strong modal to drive the feature map refinement of the weaker ones in the global and categorical perspectives by way of two sophisticated plug-and-play modules: the **G**lobal **K**nowledge-**G**uided (**GKG**) and **C**lass **K**nowledge-**G**uided (**CKG**) gated modules. The whole network then is optimized using a holistic loss function. While we were developing IKD-Net, we also established a new dataset called the **N**ational Agriculture Imagery Program and **3**D Elevation Program **C**ombined dataset in **California (N3C-California)**, which provides a particular benchmark for multi-modal joint segmentation tasks. In our experiments, IKD-Net outperformed the benchmarks and state-of-the-art methods both in the N3C-California and the small-scale ISPRS Vaihingen dataset. IKD-Net has been ranked first on the real-time leaderboard for the GRSS DFC 2018 challenge evaluation until this paper's submission.

## 1. Introduction

With the rapid development of sensors like optical cameras, radar, and 3D scanners, the era of big data has arrived; and multi-modal data for earth observation has emerged as a research frontier in remote sensing (RS), especially for land-cover semantic segmentation tasks (Ghamisi et al., 2019; Li et al., 2022; Yang et al., 2021). Multi-modal data analysis is demonstrating that it can break through the performance bottleneck of unimodal semantic segmentation by synthesizing the advantages of each data source in order to obtain more diverse feature information.

Many past studies have focused on the joint use of three-dimensional (3D) airborne LiDAR point clouds and two-dimensional (2D) satellite images. To eliminate the structure difference between the two modalities, recent researchers mostly have mapped 3D point cloud data to 2D image spaces to get products like digital surface models (DSMs) or intensity images and then extracting the 2D features for analysis and classification. CMGFNet (Hosseinpour et al., 2022) proposed a gated fusion network to achieve multi-level feature fusion between very high resolution (VHR) images and DSM. GRRNet (Huang et al., 2019) stacked the NIR-Red-Green images and the normalized DSM (nDSM) as four-channel input and utilized five gated feature labeling units to fuse the features from the encoder and decoder. MultiModNet (Liu et al., 2022) extracted features from a NIR-Red-Green image and the nDSM with the pyramid attention and the gated fusion unit, which then were joined before placing them into the decoder. Although superior to unimodal methods, these cross-modal learning methods inevitably fail to fully explore the content of each modality because the prior operation of mapping the point cloud data from 3D to 2D does some irreparable harm to important characteristics, especially the geometric structure information. To the best of our knowledge, few researchers have focused on the issue of maintaining the complete information of the raw heterogeneous data in the multi-modal land-cover semantic segmentation task.

The information imbalance phenomenon also cannot be neglected in the joint analysis and semantic segmentation of multi-modal data. We use the 3DEP-QL1 LiDAR data (3D Elevation Program LiDAR of Quality-Level-1) and NAIP (National Agriculture Imagery Program) image data to demonstrate this issue in this paper. Each tile of 3DEP point cloud data (covering an area of about 0.6 km$^2$) has a data size of about 120M and satisfies the QL1 accuracy standards (Heidemann, 2012). Specifically, the aggregate nominal pulse density (ANPD) is more than 8pls/m$^2$ (8 points per square meter) and the aggregate nominal pulse spacing (ANPS) is less than 0.35 meters. The absolute vertical accuracy (over the nonvegetated ground) is less than 0.1m and the attributes of return number and intensity values always be included. The NAIP imagery covering the same area with the above point cloud tile has a data size of only about 5M. It has a 0.6-meter ground sample distance (GSD) and affords RGB three-channel attributes. Also, the spatial resolution satisfies the standard digital orthoimage standards (Rufe, 2014). Fig. 1 illustrates the relationship of the amount of information that the two modalities provide in the joint analysis, where $x_1$ and $x_2$ stand for the modality of LiDAR and the imagery, respectively. Except for the mutual information $I(x_1, x_2)$, there may be several orders of magnitude more unique information of LiDAR than imagery, namely, $H(x_1 | x_2) \gg H(x_2 | x_1)$.

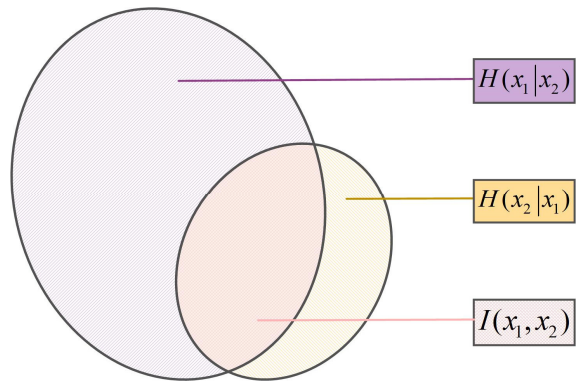

**Fig. 1.** Information diagram.

This phenomenon has attracted attention in multi-modal fusion, self-supervised learning, contrastive learning, and many other fields in the computer vision community. P4Contrast (Liu et al., 2020) leverages "pairs of point-pixel pairs" to provide extra flexibility in creating hard negatives and avoid the networks learning features only from the more discriminating one of different modalities. TupleInfoNCE (Liu et al., 2021) proposed a tuple disturbing strategy to prevent networks from largely ignoring weak modalities while only focusing on strong modalities when learning multi-modal representation. However, most of these methods concentrate on reducing the influence of imbalanced information between modalities during sample selection rather than exploiting them within network optimization.

Despite the increasing interest in multi-modal learning, multi-modal datasets in the remote sensing community are scant (Yang et al., 2021), with just two widely used datasets containing point clouds and images (i.e., the ISPRS Vaihingen semantic dataset (Rottensteiner et al., 2014) and the GRSS DFC 2018 dataset (Xu et al., 2019)). Both datasets merely cover urban land in small areas and are insufficient to quantitatively evaluate multi-modal algorithms. Thus, large-scale multi-sensor datasets are urgently needed. To fill this gap, we presented our **N**ational Agriculture Imagery Program and **3**D Elevation Program **C**ombined dataset in **California** (**N3C-California**). The LiDAR in our dataset is from 3DEP public data and the aerial imagery is from NAIP public data. We performed geometric registration and cropping on the above data and published the corresponding four categories' pixel-level labels, thus providing a quantitative evaluation of the algorithms in the field of multi-modal earth observation. The pixel-level labels were mapped from the classification attributes of the dense point clouds, which were manually annotated with very high precision, as shown in Fig. 2.

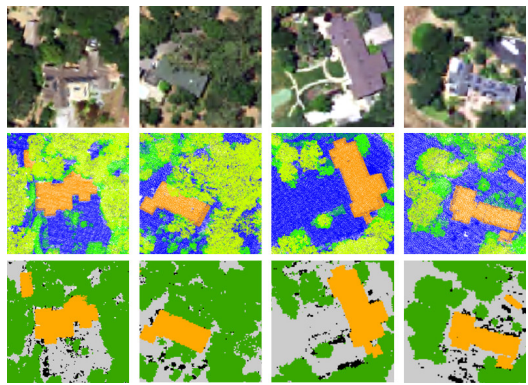

**Fig. 2.** The aerial images, labeled point clouds, and labels projected to the ground in N3C-California.

We synthesized the information discarded by previous methods in the preprocessing stage and fully exploited the inherent differences between modalities. To tackle the issues we discovered, we propose **Imbalance K**nowledge-**D**riven Multi-modal **Net**work (**IKD-Net**) to enables weak modality to get affluent information from stronger ones and promotes the modality synergy substantially.

The main contributions of this paper can be summarized as follows:

1. A specialized benchmark dataset called **N3C-California** for quantitative evaluation in multi-

modal joint segmentation tasks. **N3C-California** is the largest coverage area annotated LiDAR-imagery dataset to date.

2.    A novel efficient architecture called **IKD-Net**, which extracts features from raw multi-modal data directly rather than from their abridged derivatives. Its end-to-end disentangled dual-stream backbone helps to keep the information of heterogeneous modalities intact. The detailed ablation analysis and extensive comparative experiments in this paper on N3C-California and two other multi-modal datasets validated the design logic and superiority of IKD-Net.

3.    Two plug-and-play gated modules **G**lobal **K**nowledge-**G**uided (**GKG**) and **C**lass **K**nowledge-**G**uided (**CKG**) that take advantage of the inherent imbalance information between two RS modalities. These modules provide new insight into multi-modal data interaction.

4.    A well-designed **holistic loss function** that consists of two single-task loss functions and a pixel-wise similarity loss to maintain the balance of the parameter flow in each branch during network optimization.

## 2.    Related Work

### 2.1  Semantic segmentation for 3D LiDAR point clouds

Point clouds are essentially low-resolution resamplings of the 3D physical world. The design of learning-based semantic segmentation methods for point clouds is closely related to the data structure of 3D representations (Guo et al., 2020).

Some methods first convert 3D point clouds into intermediate regular structures and extract the features with mature 2D or 3D convolution thereafter. The segmentation results then are finally projected back to the original point clouds. These semantic segmentation methods are classified as projection-based and discrete-based. Projection-based approaches can be divided into multi-view representation (Audebert et al., 2016; Boulch et al., 2017; Lawin et al., 2017; Tatarchenko et al., 2018) and spherical representation (Iandola et al., 2016; Milioto et al., 2019; Wu et al., 2018; Wu et al., 2019) according to the projection process. Discretization-based approaches convert point clouds into discrete representations, which include dense discretization representation (Huang and You, 2016; Long et al., 2015; Meng et al., 2019; Tchapmi et al., 2017) and sparse discretization representation (Choy et al., 2019; Graham et al., 2018; Rosu et al., 2019; Su et al., 2018). These methods unfortunately fail to take full advantage of the underlying geometric and structural information as the projection step inevitably leads to missing information.

In 2017, Qi et al. proposed the pioneering work PointNet (Qi et al., 2017a), which extracts features point-by-point with a shared multilayer perceptron (MLP), opening up point-based methods whereby the network directly acts on irregular point clouds (Chen et al., 2019a; Engelmann et al., 2018; Jiang et al., 2018; Qi et al., 2017b; Vaswani et al., 2017; Yang et al., 2019; Zhang et al., 2019b; Zhao et al., 2019a; Zhao et al., 2019b). In addition to MLP networks, some studies have been devoted to specific point convolution operators (Engelmann et al., 2020; Hua et al., 2018; Thomas et al., 2019; Wang et al., 2018); methods based on recurrent neural network (RNN) (Engelmann et al., 2017; Huang et al., 2018; Liu et al., 2017; Ye et al., 2018; Zhao et al., 2019c); and graph-based methods (Landrieu and Boussaha, 2019; Ma et al., 2020; Wang et al., 2019; Wei et al., 2020; Xu and Lee, 2020; Zhiheng and Ning, 2019).

Most point cloud algorithms can handle only small ranges of point clouds, where the process of chunking may destroy the overall geometry of the point clouds. There are a few algorithms for large-scale point clouds, but they have computationally costly pre- or post-processing steps (Chen et al., 2019b; Landrieu and Simonovsky, 2018; Rethage et al., 2018). RandLA-Net (Hu et al., 2020), on the other hand, adopted a random sampling strategy to continuously downsample large-scale point clouds, which greatly reduces the computational effort and preserves the complex geometric structure in large-scale point clouds using a local feature aggregation module.

## 2.2 Multi-modality learning

According to the time point of feature fusion, multi-modal learning can be roughly divided into three categories (early, middle, and late) which correspond to data-level, feature-level, and decision-level fusion, respectively.

The early fusion strategy performs data fusion at the front end of the network and further inputs the merging layer into a single branch network for segmentation. Nahhas et al. (2018) concatenated three LiDAR-derived features (DSM, DEM, and nDSM), seven shape-derived features, three image-spectral-derived features, and eight image-texture-derived features together and then reduced the dimension by an autoencoder before using CNN to abstract the deep features. Huang et al. (2019)

put the near-infrared (NIR), red, and green bands from images and nDSM from LiDAR into a modified residual learning network and a gated feature labeling (GFL) process to extract buildings. The above early fusion strategies only treat LiDAR data as supplementary information to images, however, and ignore the discrepancies between the two modalities.

The middle fusion strategy focuses on the inter-feature combination and interaction. Fusion-FCN (Xu et al., 2019) consists of three branches that deal with the merging band of VHR image and LiDAR intensity raster data, nDSM, and high spectral data. The intermediate features are fed into a 1×1 *conv* to accomplish the final segmentation. HAFNet (Zhang et al., 2020) uses two parallel SegNet structures (Badrinarayanan et al., 2017) to extract unimodal features from image-derived RGB and LiDAR DSM data. The outputs from the counterpart layers of the two streams are processed by an attention-aware multi-modal fusion block (Att-MFBlock), which then are put into the third stream, a similar SegNet structure, to learn the multi-modal features. Zhang et al. (2017) proposed an improved FCN model which contains two parallel encoders for images and 2D elevation features from LiDAR. The feature maps from the LiDAR stream are fused with those from the image stream after every convolution module. The outputs of the two encoders and the front-end convolution modules are concatenated and fed into a decoder of the classic FCN. However, middle-level fusion is a challenging direction. Besides abstracting discriminative unimodal features, the feature-level fusion strategies should be able to distinguish the inter-modal differences and balance each modality's contribution to synthesize the high-level cross-modal features.

The late fusion strategy generally uses individual branches to extract the features of each modality, and the results are fused directly in the decision phase. Marmanis et al. (2018) proposed the Holistically-Nested Edge Detection (HED) network to fuse the boundary prediction of the separate streams. Despite having a high degree of flexibility, relatively little research has been done for late fusion because it discards cross-modal interactions and modalities cannot be adequately interrelated.

## 2.3 Attention and gating mechanism

It is commonly believed that the human eye can quickly locate the key things that are meaningful from a cluttered picture. Researchers apply this thinking in deep learning and in response have proposed the concept of attention mechanisms. Remarkable results have been achieved in natural language processing (NLP) (Galassi et al., 2020), speech recognition (Chorowski et al., 2015), and image perception (Fu et al., 2019). In 2014, a Google Mind team (Mnih et al., 2014) used an attention mechanism based on reinforcement learning (RL) in a recurrent neural network (RNN), which not only looks at the image as a whole but also extracts the necessary information from the local area. Their approach achieved excellent performance on image classification tasks and was the beginning of a trend toward widespread application of attention mechanisms. Bahdanau et al. (2014) introduced the attention mechanism concept to the NLP field for the first time. Yin et al. (2016) suggested three alternatives for employing attention mechanisms in CNNs and conducted an early exploration of their application in CNNs. Hu et al. (2018) designed the squeeze-and-excitation (SE) block to generate a channel weight distribution vector to realign the correlation between feature channels. On this basis, using the selective kernel network (SKNet), Li et al. (2019) introduced lightweight multi-channel multi-scale channel attention to obtain channel-boosted features. With the convolutional block attention module (CBAM), Woo et al. (2018) sequentially applied the attention mechanism to the input feature map in both the channel and space dimensions to produce a refined feature. The dual attention network (DANet) (Fu et al., 2019) utilized two parallel branches to produce position and channel-enhanced feature maps, which were summed and then convolved to derive the image segmentation results. In conclusion, most of the existing attention mechanism approaches can obtain the weight distribution map from a single input and then act on the input itself. These methods are limited by the inherent ceiling on the amount of information in the input itself and can only fine-tune a feature map based on the contextual relationships between the pixels within the image. Our novel approach introduces multi-modal data into the process and uses strong modal point clouds to generate a weight distribution map to "teach" the feature redistribution of weak modality images and thereby break the bottleneck of unimodal information.

## 3.   N3C-California Dataset

Despite the rapid development of earth observation methods for multi-modal data, there are unfortunately only a few LiDAR-imagery multi-modal datasets  that are dedicated to remote sensing tasks, of which ISPRS Vaihingen (Rottensteiner et al., 2014) and GRSS DFC 2018 (Xu et al., 2019) are the most commonly used. The ISPRS Vaihingen dataset provides aerial imagery in the 2D semantic labeling contest and LiDAR in the 3D semantic labeling contest. As it was not designed as a unified multi-modal benchmark, the number of categories of aerial imagery does not correspond to the number of LiDAR point clouds. The GRSS DFC 2018 dataset provides only 14 pairs of aerial imagery and LiDAR data. Both of the above datasets cover a limited range of urban areas, and the landforms are relatively simple.

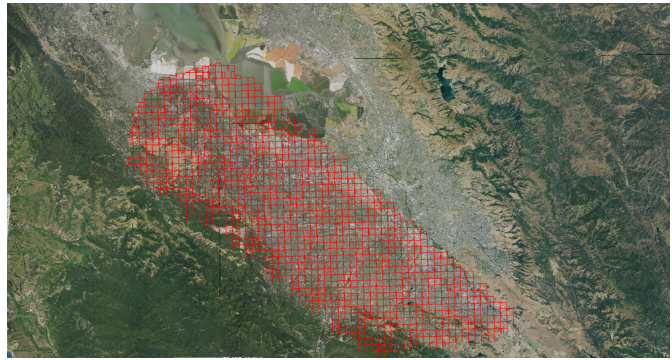

**Fig. 3.** The coverage area (red grids) of the N3C-California dataset.

To fill the gap, we introduce here our **N**ational Agriculture Imagery Program and **3**D Elevation Program **C**ombined dataset in **California** （**N3C-California**) to address the need for a benchmark specifically for multi-modal joint land-cover segmentation tasks. As shown in Fig. 3, N3C-California covers most of Santa Clara County, California and contains 1,212 pairs of LiDAR, DSM, and image tiles. The DSM is obtained by projecting the elevation of the point cloud. To facilitate the downstream tasks of remote sensing, N3C-California provides four semantic categories (ground, tree, building, and others). Fig. 4 shows several samples from N3C-California.

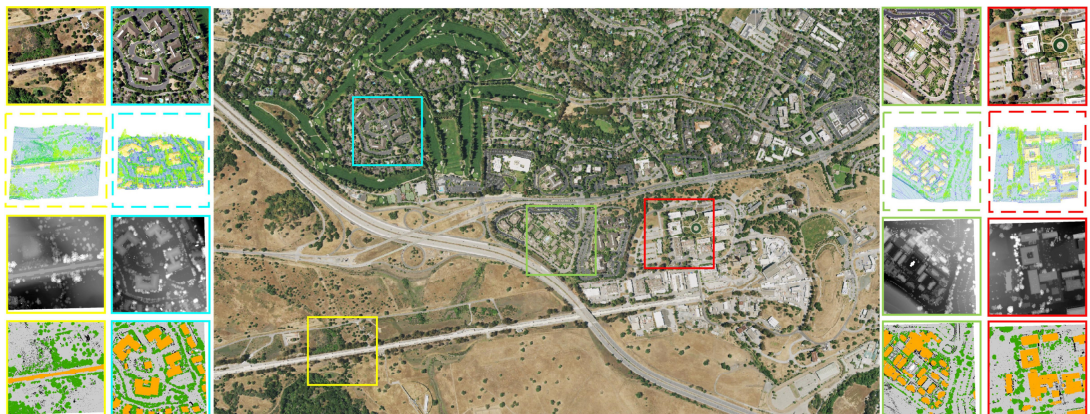

**Fig. 4.** Several samples in N3C-California, from top to bottom: aerial images, point clouds, DSM, and ground truth.

Table 1 presents an attribute comparison between N3C-California, ISPRS Vaihingen, and GRSS DFC 2018, demonstrating that N3C-California offers significant advantages in quantity and coverage. In addition to urban areas, N3C-California also includes rural regions and is hundreds of times larger than the other two datasets for total area.

**Table 1.** Attribute comparison of N3C-California, ISPRS Vaihingen, and GRSS DFC 2018.

|  | N3C-California | ISPRS Vaihingen | GRSS DFC 2018 |
|---|---|---|---|
| Number of tiles | 1212 | 33 | 14 |
| LiDAR ANPD (pls/m$^2$) | ≥8 | 4 | 10 |
| Image dimension (px) | 1304×1304 (avg) | 2493×2063 (avg) | 11920×12020 |
| GSD (cm/pixel) | 100 | 9 | 5 |
| Coverage (km$^2$) | 725.72(urban&rural) | 1.36(urban) | 5.01(urban) |
| Classes | 4 | 6(image)/9(LiDAR) | 20 |

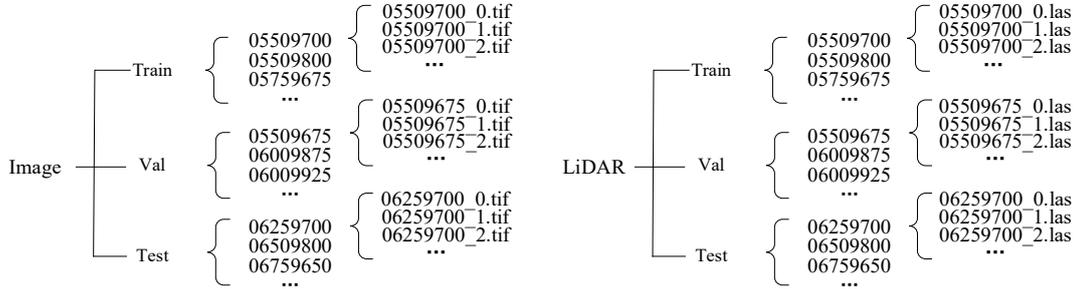

**Fig. 5.** The organization of the cropped data of N3C-California.

For the convenience of model training, we cropped the data into 10,800 image patches with 512×512 pixels of 20% overlaps. Fig. 5 shows the organization of the cropped data of N3C-California, in which the numbers without and with extension represent the names of tile data and patch data respectively. The training set, validation set, and test set were randomly divided according to the ratio of 8:1:1, as illustrated in Fig. 6. By contrast, the division of ISPRS Vaihingen (11 samples for training, five samples for validation, and 17 samples for testing) and GRSS DFC 2018 (four samples for training, none for validation, and 10 samples for testing) do not exactly correspond to the general setting of deep network training, as they are not specifically multi-modal deep benchmarks.

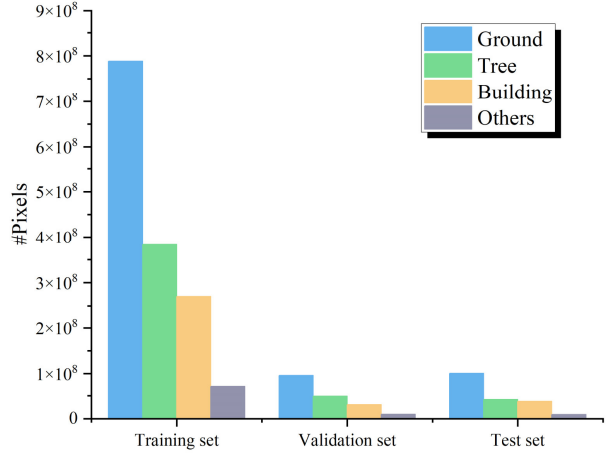

**Fig. 6.** Number of annotated pixels in N3C-California.

## 4. Methodology

### 4.1 Heterogeneous network

As point clouds and images belong to different dimensional spaces, the common strategy of combining the abridged 2D feature maps from LiDAR with images before insertion in the network will sacrifice the rich information of the former and fail to exploit the multi-modal features fully. To this end, we propose our novel **I**mbalance **K**nowledge-**D**riven Multi-modal **Net**work (**IKD-Net**), which can extract heterogeneous features from LiDAR point clouds and image data in parallel. Fig. 7 presents the workflow of the disentangled dual-stream heterogeneous network. The two branches were crafted to be similar encoder-decoder structures, thereby making it possible to obtain similar size feature maps at the same branch depth. The LiDAR stream utilizes RandLA-Net (Hu et al., 2020) as the backbone network to extract 3D features, while the image stream utilizes UNet (Ronneberger et al., 2015) for 2D feature extraction. Both branches can access individual knowledge, such as geometry in 3D space for LiDAR and texture and color information in 2D space for images. Unlike previous approaches that treat features from different modalities as homogeneous and design symmetric feature interaction modules, we exploit the affluent knowledge of LiDAR (strong modality) to drive the refinement of feature maps from images (weaker modality) with our **G**lobal **K**nowledge-**G**uided (**GKG**) gated module and **C**lass **K**nowledge-**G**uided (**CKG**) gated module in the decoder parts. Four GKG gated modules obtain the global feature distribution from the LiDAR features at different resolutions to guide the image features at the same network depth to focus on the region of interest (ROI). The CKG gated module, which is applied at the end of the dual-stream architecture, provides the performance evaluation of each category from a global perspective and achieves the coarse-to-fine segmentation. Before the feature interactions, a front-end projection transformation module called the **D**imension **S**ensor (**DS**) is performed.

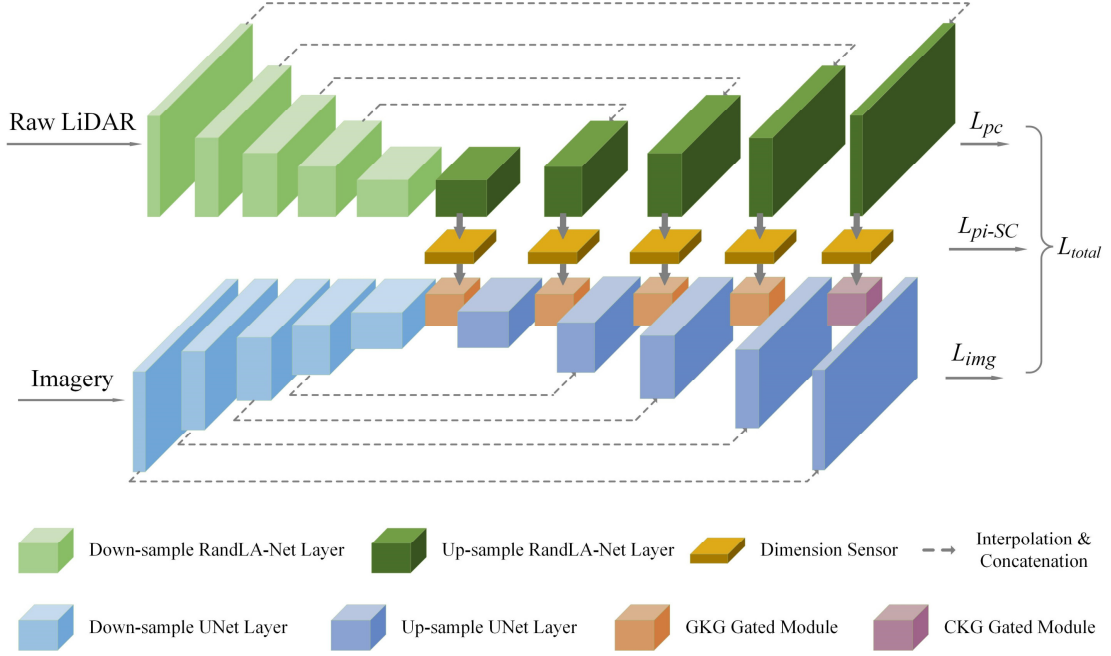

**Fig. 7.** The workflow of IKD-Net.

## 4.2 Dual-stream feature extraction

While the dual-stream network design aims to fully preserve the information of unimodal data and effectively leverage multi-modal features to facilitate feature interaction, the dual branches at the same time must obtain feature maps of the same size at the same depth of the network. If the input image size is 512×512, the number of input LiDAR points should be on the order of $2 \times 10^5$. Therefore, the 3D branch must have an exceptional ability to handle large-scale point clouds. To meet this need, we selected RandLA-Net as the 3D backbone and designed the corresponding encoder-decoder 2D network. In the encoder part, RandLA-Net first uses a linear transformation layer to expand the feature dimension to 8. The specially-designed local feature aggregation (LFA) and random sampling modules are repeated in the subsequent four downsampling layers. The LFA module consists of two crucial blocks (local spatial encoding and attention pooling).

Local spatial encoding embeds the local geometric pattern for each individual point. Specifically, it first finds the K nearest points around each point with the *k*-nearest neighbors (*k*-NN) algorithm. Then, within the above K points, the information of the coordinates is aggregated. In this paper, K is set to 16. The aggregated calculation formula is as follows:

$$r_i^k = MLP\left(p_i \oplus p_i^k \oplus \left(p_i - p_i^k\right) \oplus \left\| p_i - p_i^k \right\|\right) \tag{1}$$

where $r_i^k$ represents the coordinate encoding value of the *i*-th point and its *k*-th neighbor point. *MLP* denotes the multilayer perceptron. The four terms in the outermost bracket are the coordinates of the center point, the coordinates of its *k*-th neighbor point, the relative coordinate difference, and the relative distance between the two points. $\oplus$ stands for concatenation.

Secondarily, the transformed encoding $r_i^k$ is concatenated with the corresponding feature $f_i^k$:

$$\widetilde{f}_i^k = r_i^k \oplus f_i^k \tag{2}$$

The attention pooling module further refines the feature encodings obtained in the previous step $\widetilde{F}_i = \{\widetilde{f}_i^1 \cdots \widetilde{f}_i^k \cdots \widetilde{f}_i^K\}$ (i.e., the feature maps are multiplied at the pixel level with the weight distribution maps generated on them). Ultimately, the results are accumulated to obtain the aggregated features $\widehat{f}_i$ of the individual points.

We use features from the strong modal point clouds in the decoder to guide the images for feature map refinement. The refined images then use interpolation and convolution blocks to restore the image size, and the LiDAR stream does likewise.

## 4.3 Dimension sensor

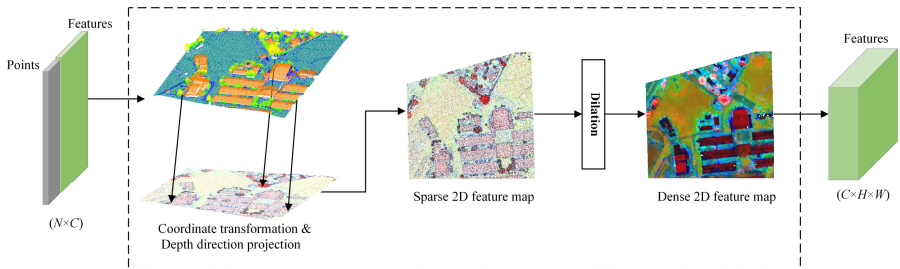

**Fig. 8.** Dimension sensor.

We designed our plug-in DS module for the purpose of aligning the high-dimensional features to the low-dimensional ones, as shown in Fig. 8. The dashed box contains an example of a case where the number of channels is 1. First, we performed coordinate transformation and depth direction projection on each 3D point to produce a feature map in the same metric space as the image. However, there are bound to be pixels not covered by points in feature maps (i.e., hole pixels), which introduce significant inaccuracies when the images are superimposed. Thus, we executed a dilation operation with the trick of max pooling, which fills the hole pixels without increasing the complexity of the network, as depicted in Fig. 9. After passing through the DS, the point cloud features can be converted to the image space with a high degree of fit.

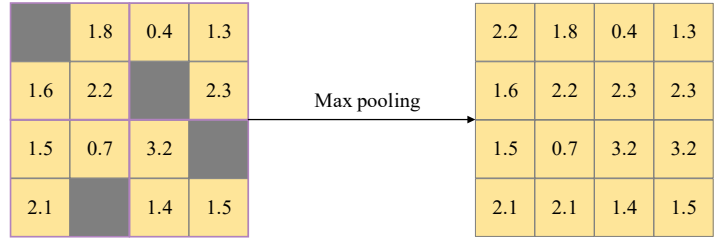

**Fig. 9.** Dilation operation.

## 4.4 Global knowledge-guided gated module

Owing to the nature of spotlighting the neighborhoods of convolutional kernels, the information flow in convolutional neural networks is restricted to local areas (Zhao et al., 2018). A spatial attention mechanism can generate an overall probability map to focus on the ROI, thereby extending the global contextual understanding of complex scenes. The probability distribution of the existing methods is derived from the input itself, but due to the limitation of unimodal data information capacity, there is a ceiling to this refinement. To tackle this problem, we proposed a GKG gated module to provide a global probability distribution map utilizing feature maps from strong modality point clouds to guide the further refinement of the image feature map. The structure of our GKG gated module is shown in Fig. 10.

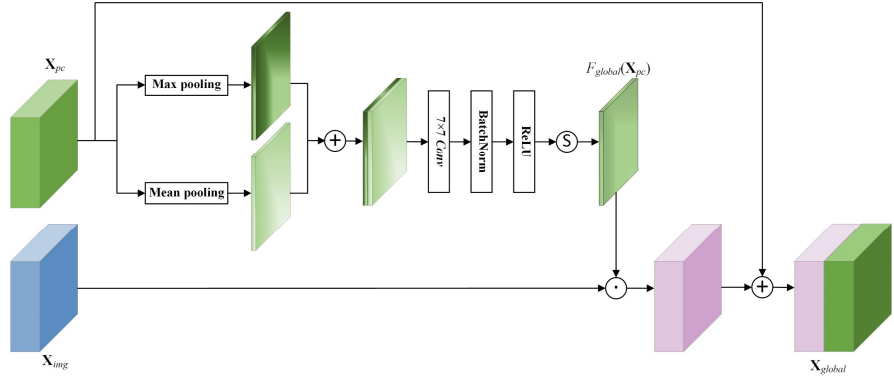

**Fig. 10.** The structure of the GKG gated module. +, · and S represent concatenation, multiplication along the channel, and sigmoid function, respectively.

The global attention map from point cloud $F_{global}(\mathbf{X}_{pc})$ is defined as:

$$F_{global}\left(\mathbf{X}_{pc}\right) = \sigma\left(g^{7\times7}\left(\left[AvgPool(\mathbf{X}_{pc}), MaxPool(\mathbf{X}_{pc})\right]\right)\right) \qquad (3)$$

The avg- and max-pooling operations generate compact feature representations in the spatial dimension. $g^{7\times7}$ is a sequence operation of 7×7 *conv*, batch normalization, and ReLU. $\sigma$ denotes the sigmoid function. By this sequential operation, the feature map of the strong modal point cloud is compressed into a spatial-wise weight distribution map.

Then, $F_{global}(\mathbf{X}_{pc})$ is applied to drive the attention boosting of the weak modalities. The spatial attention-boosted feature map is obtained from the image. Finally, the spatially enhanced image feature maps are concatenated with the point cloud feature maps. In summary, the output of the GKG gated module is:

$$\mathbf{X}_{global} = \left[F_{global}\left(\mathbf{X}_{pc}\right) \odot \mathbf{X}_{img}, \mathbf{X}_{pc}\right] \qquad (4)$$

where $\odot$ represents multiplication along the channel.

The GKG gated module leverages the higher-level semantic information of the strong modality to provide guidance on the global distribution for the weak modality, refining the latter's understanding of the global context.

## 4.5 Class knowledge-guided gated module

Besides global information, inter-class variability also plays an influential role in segmentation tasks. Assuming that there is a fixed-length encoding for each category (i.e., the theoretical class center), all the pixels in the global scene belonging to that category should make a contribution (Zhang et al., 2019a). Further, these encodings can in turn optimize the category attribution of each pixel in the scene. Accordingly, we added a coarse-to-fine structure called the CKG gated module at the output part of IKD-Net, as shown in Fig. 11.

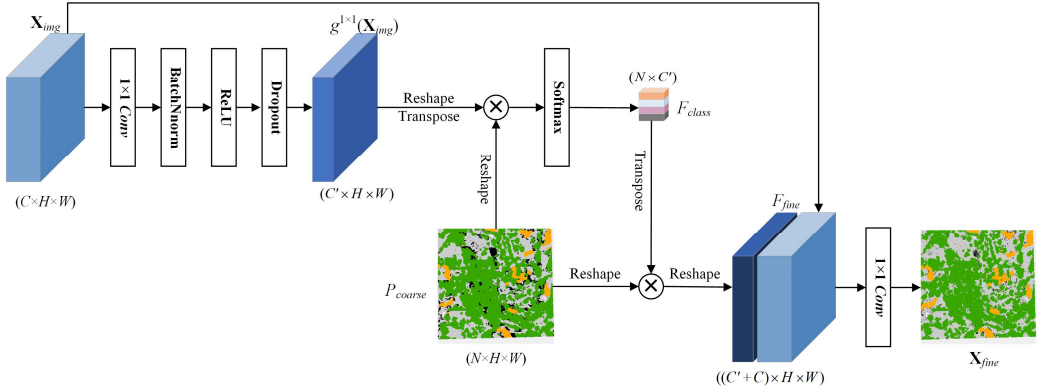

**Fig. 11.** The structure of the CKG gated module. × stands for the matrix multiplication.

First, the CKG gated module distills the contextual information and generates the category probability map $F_{class}$ with the coarse segmentation result $P_{coarse}$ from the point clouds and the feature map $\mathbf{X}_{img}$ of the corresponding image. Each row of $F_{class}$ provides the performance evaluation of a category from a global perspective by converging the feature vectors of all the pixels belonging to the category.

In detail, we put $\mathbf{X}_{img}$ through a sequence operation $g^{1\times1}$ of 1×1 *conv*, batch normalization, and ReLU to reduce the channel dimension from $C$ to $C'$. Then, a reshape operation is applied to $g^{1\times1}(\mathbf{X}_{img})$ and $P_{coarse}$ respectively:

$$g^{1\times1}\left(\mathbf{X}_{img}\right) \in \mathbb{R}^{C'\times H\times W} \rightarrow g^{1\times1}\left(\mathbf{X}_{img}\right) \in \mathbb{R}^{C'\times HW} \tag{5}$$

$$P_{coarse} \in \mathbb{R}^{N\times H\times W} \rightarrow P_{coarse} \in \mathbb{R}^{N\times HW} \tag{6}$$

where $N$ is the category number.

The category probability map $F_{class} \in \mathbb{R}^{N\times C'}$ is calculated as:

$$F_{class} = \mathrm{softmax}\left(P_{coarse}\left(g^{1\times1}\left(\mathbf{X}_{img}\right)\right)^{\mathrm{T}}\right) \tag{7}$$

Second, taking the coarse class distribution of each pixel as the mediator, the attentional class feature vector of each pixel is obtained by multiplying the coarse segmentation result $P_{coarse}$ and the category probability map $F_{class}$. The attentional class feature map is:

$$F_{fine} = P_{coarse}^{\mathrm{T}} F_{class} \tag{8}$$

For the subsequent operation, $F_{fine}$ is reshaped:

$$F_{fine} \in \mathbb{R}^{C' \times HW} \rightarrow F_{fine} \in \mathbb{R}^{C' \times H \times W} \tag{9}$$

Finally, the concatenation of $F_{fine}$ and $\mathbf{X}_{img}$ is put into a 1×1 $conv\, f^{1 \times 1}$ to obtain the class boosting feature map:

$$\mathbf{X}_{fine} = f^{1 \times 1}[\mathbf{X}_{img}, F_{fine}(P_{coarse}, \mathbf{X}_{img})] \tag{10}$$

The CKG gated module guides each pixel in the high-level feature maps of an image (weak modality) to adaptively approach the theoretical class centers according to the segmentation results of the point clouds (strong modality).

## 4.6  Loss function

In order to maintain the balance of the parameter flow in each branch during network optimization, we proposed a holistic loss function, which consists of three types of supervision: two single-task loss functions and a pixel-wise similarity loss.

**Holistic Loss.** Serving as the whole objective function, the holistic loss enables the network to be trained in an end-to-end manner, which is summarized as:

$$L_{total} = L_{CE}(P(x_{img})) + L_{CE}(P(x_{pc})) + L_{pi-SC}(P(x_{img}), P(x_{pc})) \tag{11}$$

where $x_{img}$ and $x_{pc}$ denotes the input image and LiDAR, respectively; $P(*)$ represents the final probability distribution map; $L_{CE}$ is the segmentation cross-entropy loss; and $L_{pi\text{-}SC}$ is the pixel-wise similarity loss.

**Single-task Loss.** For each branch of semantic segmentation, given predict $P(x)$ and ground truth $y$, we use cross-entropy loss to optimize, which is as follows:

$$L_{CE}(P(x)) = -\sum_{i=1}^{N}\sum_{c=1}^{M} y_{ic} \log(P(x_{ic})) \tag{12}$$

where $N$ represents the number of pixels or points and $M$ denotes the number of categories.

**Pixel-wise Similarity Constraint.** To make the convergence spaces of the images and point clouds of the same scene as close as possible, we straightforwardly add similarity constraints.

Further, inspired by knowledge distillation (Hinton et al., 2015; Liu et al., 2019), we consider the class distribution of the point clouds (strong modality) $P(x_{pc})$ as soft targets to guide the images (weak modality) $P(x_{img})$, thus improving the accuracy of 2D semantic segmentation.

We use Kullback-Leibler divergence to implement this pixel-wise similarity constraint, which is formulated as follows:

$$L_{pi-SC}(P(x_{img}), P(x_{pc})) = \frac{1}{W \times H} \sum_{i=1}^{W \times H} \text{KL}(P_i(x_{img}) \| P_i(x_{pc})) \qquad (13)$$

where

$$\text{KL}(P_i(x_{img}) \| P_i(x_{pc})) = P_i(x_{img}) \log(\frac{P_i(x_{img})}{P_i(x_{pc})}) \qquad (14)$$

## 5. Experiments

In this section, we explain the experimental setups and evaluation metrics. We conducted sufficient ablation studies to verify the rationality of our sophisticated modules and the overall structure of IKD-Net. We visualize the outcomes of each module here. Then, we compare our IKD-Net results with those of the state-of-the-art (SOTA) methods on the N3C-California, ISPRS Vaihingen, and GRSS DFC 2018 datasets. Finally, we demonstrate how our method achieved outstanding performance on all three datasets.

### 5.1 Experimental setting

#### 5.1.1 Implementation details

All the experiments were conduct on a Linux PC equipped with an NVIDIA GeForce RTX 3090 24G GPU. The code of our own architecture and the code we reproduced are based on the PyTorch deep learning framework. During training, the batch size was set to 2 for the experiments on all the datasets. Each epoch had 1,000 iterations, and the maximum number of epochs was always 50. On the N3C-California dataset, the Adam algorithm with a 0.001 learning rate was employed for optimization. For the other datasets, the SGD method with a 0.01 learning rate, 0.0001 weight decay, and 0.9 momentum was chosen. The input images were 512×512 pixel in size. A total of 131,072 randomly selected points with x, y, z, intensity attributes and the return number attributes from the appropriate LiDAR patch covering the same area were fed into the networks simultaneously.

### 5.1.2 Evaluation metrics

The results were evaluated by overall accuracy (OA), mean accuracy (Mean Acc), Cohen's Kappa (Kappa), and mean intersection over union (mIoU).

OA is defined as the ratio of the number of correctly classified pixels $p_{correct}$ to the total number of pixels $p_{all}$.

$$OA = \frac{p_{correct}}{p_{all}} \tag{15}$$

OA is simple to calculate but is easily dominated by a large number of samples in the case of unbalanced samples, which can be addressed by three other metrics.

We assume $T_P^k$, $F_P^k$, $T_N^k$, $F_N^k$ represent the true positive number, the false positive number, the true negative number, and the false negative number for $k$-th class, respectively, in the confusion matrix. Accuracy and IoU for $k$-th class ($Acc_k$ and $IoU_k$) are defined as:

$$Acc_k = \frac{T_P^k}{T_P^k + F_N^k} \tag{16}$$
$$\tag{17}$$

$$IoU_k = \frac{T_P^k}{T_P^k + F_P^k + F_N^k} \tag{18}$$

For total K categories, Mean Acc and mIoU are defined as:

$$Mean\ Acc = \frac{1}{K}\sum_{k=1}^{K}\frac{T_P^k}{T_P^k + F_N^k} \tag{19}$$

$$mIoU = \frac{1}{K}\sum_{k=1}^{K}\frac{T_P^k}{T_P^k + F_P^k + F_N^k} \tag{20}$$

The formula for Kappa is:

$$Kappa = \frac{OA - p_e}{1 - p_e} \tag{21}$$

where

$$p_e = \frac{\sum_{k=1}^{K}\left(T_P^k + F_P^k\right)\left(T_P^k + F_N^k\right)}{p_{all}^2} \tag{22}$$

## 5.2 Results on N3C-California dataset

In the ablation studies section, we marked the dual-stream backbone of our IKD-Net as IKD-Net-, which indicates that we discarded the GKG and CKG gated modules and it now was equipped only with one single-task loss function (the image segmentation loss function).

### 5.2.1 Ablation study for GKG gated module

The four structured GKG gated modules in the two-branch structure provided multi-resolution weight redistribution maps for top-to-down weak modality feature map refinement under strong modality guidance. To demonstrate the effect of the structured GKG modules, we gradually increased the number of GKG modules on the backbone IKD-Net-. The experimental results are shown in Table 2. In rows 2-5, we replaced the GKG gated modules with the simple feature concatenation operations at the same positions.

**Table 2.** Ablation study for different number of GKG gated modules. CAT represents the simple feature concatenation operations.

| Backbone | #GKG | #CAT | OA | Mean Acc | Kappa | mIoU |
|----------|------|------|--------|----------|--------|--------|
| UNet | - | - | 0.8635 | 0.6717 | 0.7722 | 0.5943 |
| IKD-Net- | - | 1 | 0.8707 | 0.8487 | 0.7917 | 0.6345 |
| IKD-Net- | - | 2 | 0.9059 | 0.8702 | 0.8446 | 0.6838 |
| IKD-Net- | - | 3 | 0.9167 | 0.8920 | 0.8638 | 0.7081 |
| IKD-Net- | - | 4 | 0.9218 | 0.9077 | 0.8730 | 0.7197 |
| IKD-Net- | 1 | - | 0.8825 | 0.8563 | 0.8092 | 0.6533 |
| IKD-Net- | 2 | - | 0.9100 | 0.8806 | 0.8525 | 0.6949 |
| IKD-Net- | 3 | - | 0.9170 | 0.8949 | 0.8642 | 0.7087 |
| IKD-Net- | 4 | - | **0.9275** | **0.9142** | **0.8822** | **0.7289** |

At least four primary conclusions can be drawn from Table 2. First, as implied in rows 1-5, the supplemental feature maps from the LiDAR stream greatly improved the image segmentation, and the more information that was provided from the former the greater the accuracy improvement. Simply overlaying four multi-resolution feature maps from the LiDAR stream (row 5), our IKD-Net- backbone outperformed UNet by nearly 0.06 in OA, over 0.23 in Mean Acc, over 0.1 in Kappa, and over 0.12 in mIoU. Second, upgrading the simple concatenation operations with GKG gated modules (row 2-5 vs. row 6-9) further enhanced the ability to jointly exploit the multi-modal features by driving the refinement of the feature distribution of the weak modality with the affluent knowledge from the strong modality. Third, as we gradually increased the number of GKGs, the accuracy steadily improved, indicating that the effects of our GKGs were cumulative. Eventually, IKD-Net- equipped with four structured stacked GKG gated modules (row 9) surpassed the baseline by over 0.06 in OA, over 0.24 in Mean Acc, 0.11 in Kappa, and over 0.13 in mIoU.

### 5.2.2 Ablation study for CKG gated module

The CKG gated module simultaneously distilled the contextual information of the strong and weak modalities to obtain the category-wise feature map, the so-called class centers. The class centers were then exploited to guide the refinement of the feature maps of the weak modal images from coarse to fine. We observed the effect of adding the CKG gated module based on the optimal structure in the last section (marked as IKD-Net- (GKG-4)), as indicated in Table 3.

**Table 3.** Ablation study for CKG gated module.

| Backbone | CKG | Loss | OA | Mean Acc | Kappa | mIoU |
|---|---|---|---|---|---|---|
| IKD-Net- (GKG-4) | - | $L_{img}$ | 0.9275 | **0.9142** | 0.8822 | 0.7289 |
| IKD-Net- (GKG-4) | ✓ | $L_{img}$ | 0.9235 | 0.9075 | 0.8754 | 0.7207 |
| IKD-Net- (GKG-4) | ✓ | $L_{img}+L_{pc}$ | **0.9319** | 0.9136 | **0.8886** | **0.7347** |

Row 2 in Table 3 contains the results of IKD-Net- (GKG-4) with the addition of the CKG gated module. The accuracy decreased slightly with respect to simple IKD-Net- (GKG-4) (row 1) on all the metrics. This decrease may have been due to CKG depending largely on the coarse segmentation process, which is not fully optimized by the single image segmentation loss function because the backward-propagation route is too circuitous for the LiDAR stream. Therefore, we incorporated an additional cross-entropy loss function to the LiDAR stream, as shown in row 3. By adding the CKG gated module to IKD-Net- (GKG-4), it eventually exceeded its counterpart without CKG on three metrics.

### 5.2.3 Ablation study for loss function

The holistic loss function $L_{total}$ takes into account both the independent optimizations of each branch and the synergy between them. Table 4 indicates the superposition effect of the three terms in $L_{total}$, which are the single image segmentation loss function, single point cloud segmentation loss function, and pixel-wise similarity constraint.

**Table 4.** Ablation study for different loss functions.

| Backbone | CKG | Loss | OA | Mean Acc | Kappa | mIoU |
|---|---|---|---|---|---|---|
| IKD-Net- (GKG-4) | ✓ | $L_{img}$ | 0.9235 | 0.9075 | 0.8754 | 0.7207 |
| IKD-Net- (GKG-4) | ✓ | $L_{img}+L_{pc}$ | 0.9319 | **0.9136** | 0.8886 | 0.7347 |
| IKD-Net- (GKG-4) | ✓ | $L_{total}$ | **0.9381** | 0.9061 | **0.8977** | **0.7550** |

As explained in the last section, the joint use of two single-task loss functions facilitated the performance of the CKC module. Furthermore, the addition of a pixel-wise similarity constraint further promoted the improvement of image segmentation accuracy, thereby surpassing its counterpart with a single image segmentation loss of nearly 0.015 in OA, over 0.02 in Kappa, and over 0.034 in mIoU.

### 5.2.4 Visualization results of GKG and CKG gated modules

In order to qualitatively analyze the effect of each module on the features, we visualized the feature maps after applying each module, as shown in Fig. 12. We took the mean value in the channel direction for the high-dimensional feature maps to normalize and stretch the obtained single-channel 2D feature map to 0-255. Finally, we performed a pseudo-color transformation to obtain the feature maps that facilitated visual interpretation.

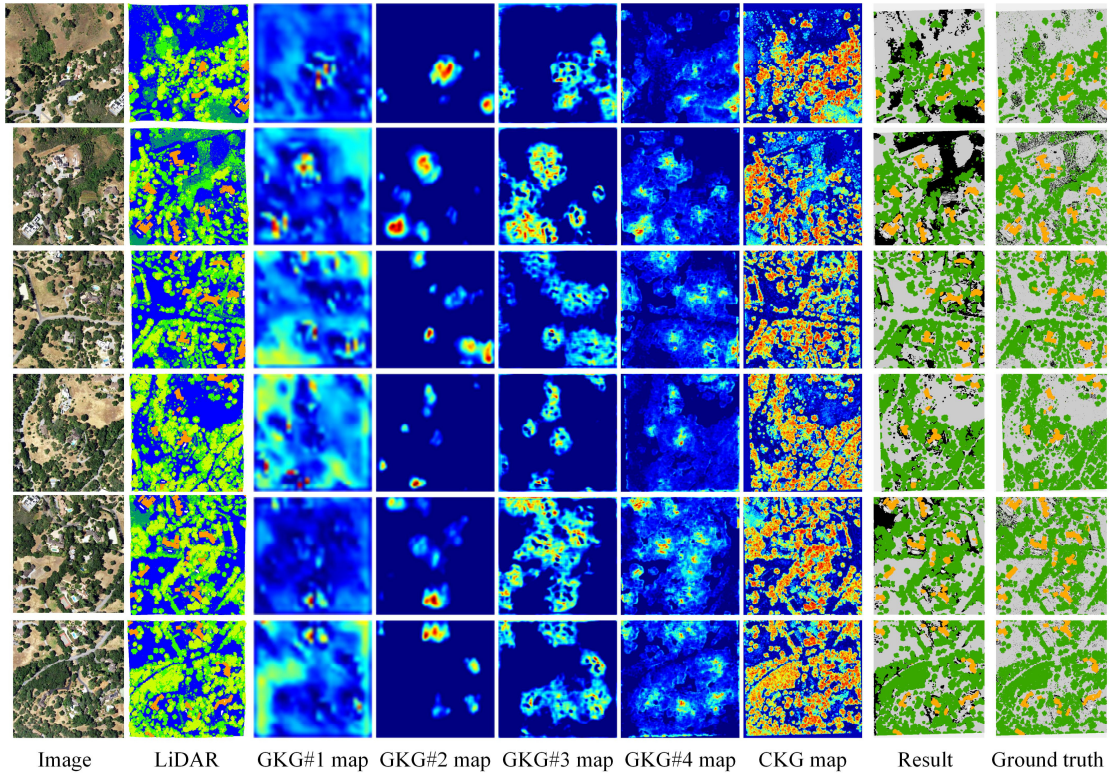

Image LiDAR GKG#1 map GKG#2 map GKG#3 map GKG#4 map CKG map Result Ground truth

**Fig. 12.** Visualization of the feature maps after applying each module.

As is evident from each row of Fig. 12, the segmentation results were progressively detailed after each module was applied. We observed that as the number of GKG modules increased, their effect of fusing global information became more pronounced, mainly in the Building category. Intuitively, in terms of the global scene, the characteristics of the buildings became more distinct compared to the ground and trees. Columns 3-6 show that the scattered buildings gradually became more distinguishable from the other categories as the boundaries became increasingly more precise. After the fourth GKG module (column 6) was applied, a large area of trees now has clear demarcation lines from the ground. The CKG module performed class-weighted refinement on the feature maps, which improved the inter-class difference and intra-class similarity, as illustrated in column 7. It further subdivided the large areas that were misclassified into the same type by the previous modules, and the boundaries between the trees (orange) and buildings (turquoise) also became sharper.

### 5.2.5 Comparing with SOTA methods

Finally, we compared the complete IKD-Net with the baseline method (UNet (Ronneberger et al., 2015), RandLA-Net(Hu et al., 2020)), the multi-modal benchmark method (Hybri-UNet (Sherrah, 2016)), and the SOTA multi-modal segmentation network in RS field (vFuseNet (Audebert et al., 2018), MultifilterCNN (Sun et al., 2018), and MFNet (Sun et al., 2021)). The same experimental hyperparameters were used for all the methods. The results are shown in Table 5.

**Table 5.** Quantitative comparison of IKD-Net and the SOTA multi-modal models.

| Method | Input | OA | Mean Acc | Kappa | IoU | | | | |
|---|---|---|---|---|---|---|---|---|---|
| | | | | | Others | Ground | Tree | Building | Mean |
| UNet (baseline) | RGB | 0.8635 | 0.6717 | 0.7722 | 0.0246 | 0.8092 | 0.7333 | 0.8101 | 0.5943 |
| RandLA-Net (baseline) | LiDAR | 0.8749 | 0.8578 | 0.8191 | 0.4006 | 0.8832 | **0.8419** | 0.6998 | 0.7064 |
| Hybri-UNet | RGB+DSM | 0.8900 | 0.7176 | 0.8166 | 0.1298 | 0.8586 | 0.7703 | 0.8713 | 0.6575 |
| vFuseNet | RGB+DSM | 0.8611 | 0.7547 | 0.7468 | **0.4973** | 0.8199 | 0.5758 | 0.7743 | 0.6668 |
| MultifilterCNN | RGB+DSM+intensity +number returns+DoG | 0.8897 | 0.7644 | 0.8168 | 0.2577 | 0.8433 | 0.7763 | 0.8263 | 0.6759 |
| MFNet | RGB+DSM +Slope angle+DoG | 0.9100 | 0.7485 | 0.8572 | 0.1429 | 0.8736 | 0.8209 | 0.8987 | 0.6840 |
| IKD-Net (ours) | RGB+LiDAR | **0.9381** | **0.9061** | **0.8977** | 0.3068 | **0.9311** | 0.8235 | **0.9587** | **0.7550** |

For each method in Table 5, the IoU of the Others category was much lower than the other three categories, which was caused by two factors. First, the Others category contained three subcategories (low vegetation, water, and road surface), which made it more difficult to obtain a unified feature description. Second, the number of pixels belonging to the Others category was only about five percent of the other classes, making it very difficult for networks to learn the discriminative features.

All the multi-modal methods exceeded the baseline method UNet in IoU, indicating that introducing the other modality indeed improved the effect of 2D semantic segmentation. However, none of them achieved a higher IoU than the baseline method RandLA-Net, which may have been due to their inevitable information loss when mapping the 3D point cloud data to the 2D image space in the preprocessing stage.

Our IKD-Net significantly surpassed the current SOTA multi-modal segmentation methods in all the metrics. IKD-Net boosted the previous optimal performance (MFNet) by more than 0.028, 0.157, 0.04, and 0.07 on OA, Mean Acc, Kappa and mIoU, respectively. In particular, our IoU in the Building category reached over 0.95, laying a better foundation for the downstream RS tasks. The excellent outcome of IKD-Net mainly can be attributed to its heterogeneous networks and well-designed feature interaction module that directly extract features from the raw data source and utilize the imbalance information between them. It is worth noting that the 2D products from 3D point cloud data (DSM, intensity image, etc.) are generated from dense point clouds with approximately $2 \times 10^6$ points in every LiDAR patch while our IKD-Net randomly selected only 131,072 points from each LiDAR patch for a compromise with the computer memory. Nevertheless, even the relatively sparse point clouds still provided a powerful knowledge-driven effect for the images, dramatically improving the segmentation accuracy.

Fig. 13 displays the qualitative comparison results on six scenes. The unimodal method, Unet, barely distinguished the Others category, which revealed that multi-modal data offers a distinct advantage in the segmentation of the more ambiguous categories. For the second scene, it is evident that IKD-Net outperformed the other SOTA multi-modal strategies in terms of completeness and edge conformity for building segmentation by correctly outlining the edges of the two connected buildings on the right side of the third scene and restoring their connected form. Although the edges of the Tree category in all six scenes were very irregular and had many scattered small areas, IKD-Net outstandingly reconstructed its rough boundary lines.

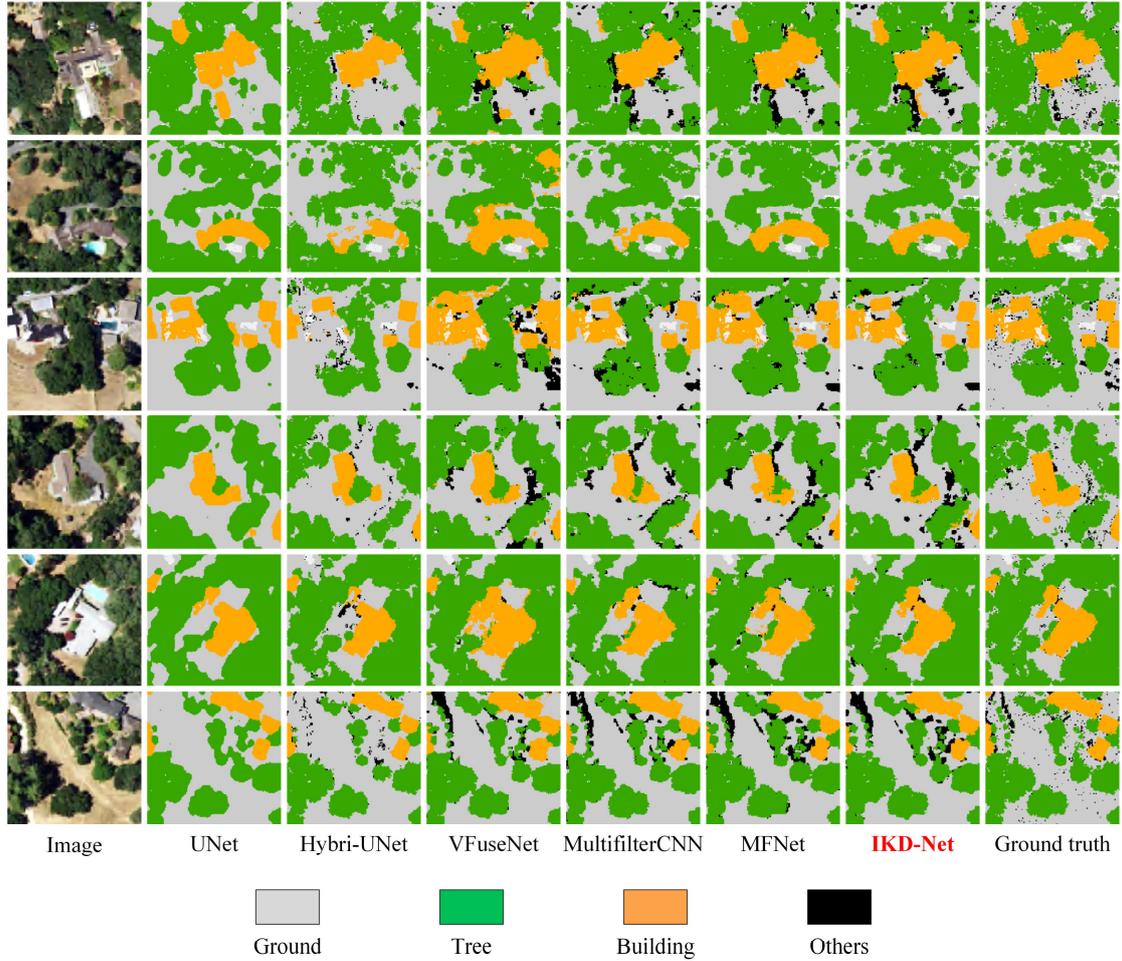

| Image | UNet | Hybri-UNet | VFuseNet | MultifilterCNN | MFNet | **IKD-Net** | Ground truth |

Ground     Tree     Building     Others

**Fig. 13.** Qualitative comparison of the baseline, the multi-modal benchmarks, the SOTA multi-modal segmentation networks, and IKD-Net (ours) on N3C-California.

## 5.3 Results on ISPRS Vaihingen dataset

As shown in Table 6, we compared our IKD-Net with the benchmark competitors using the ISPRS Vaihingen dataset. The benchmark competitors on the challenge evaluation website were measured only on their OA and F1 score rounded to three decimal places so we indicate the same criteria in Table 6. The F1 score is the harmonic mean of OA and the recall rates. The best value under a certain metric is bolded in Table 6.

Our IKD-Net ranked first among all the excellent methods on the OA and the mean F1 and achieved the best F1 score in four of the five categories. We believe the superiority of IKD-Net is due to its ability to treat the two modalities distinctly and then leverage the strong modality to drive the feature learning of the weak modality.

**Table 6.** Quantitative comparison of IKD-Net with the benchmark competitors on ISPRS Vaihingen dataset.

| Method | OA | F1 Score | | | | | |
|---|---|---|---|---|---|---|---|
| | | Imp surf | Building | Low veg | Tree | Car | Mean |
| SVL_3 (Gerke, 2014) | 0.848 | 0.866 | 0.910 | 0.770 | 0.850 | 0.556 | 0.790 |
| HUST(Quang et al.) | 0.859 | 0.869 | 0.920 | 0.783 | 0.869 | 0.290 | 0.746 |
| RIT(Piramanayagam et al., 2016) | 0.863 | 0.881 | 0.930 | 0.805 | 0.872 | 0.419 | 0.781 |
| UOA(Lin et al., 2016) | 0.876 | 0.898 | 0.921 | 0.804 | 0.882 | 0.820 | 0.865 |
| ADL_3 (Paisitkriangkrai et al., 2015) | 0.880 | 0.895 | 0.932 | 0.823 | 0.882 | 0.633 | 0.833 |
| DST_1(Sherrah, 2016) | 0.887 | 0.903 | 0.935 | 0.825 | 0.888 | 0.739 | 0.858 |
| DLR_8 (Marmanis et al., 2018) | 0.892 | 0.904 | 0.936 | 0.839 | 0.897 | 0.769 | 0.869 |
| UFMG_4 (Nogueira et al., 2019) | 0.894 | 0.911 | 0.945 | 0.829 | 0.888 | 0.813 | 0.877 |
| ONE_7 (Audebert et al., 2016) | 0.898 | 0.910 | 0.945 | 0.844 | 0.899 | 0.778 | 0.875 |
| CASIA2 (Liu et al., 2018) | 0.911 | 0.932 | **0.960** | 0.847 | 0.899 | 0.867 | 0.901 |
| IKD-Net (ours) | **0.921** | **0.961** | 0.905 | **0.872** | **0.920** | **0.925** | **0.916** |

Fig. 14 displays the qualitative results of IKD-Net and five excellent benchmark methods on the ISPRS Vaihingen dataset. Fig. 14 shows that all the methods delivered exceptional performance, while our IKD-Net excelled in integrity and accurately identified the boundaries of buildings. Moreover, only IKD-Net was able to separate the tree objects in the lower left corner of the third scene while the other methods joined them together.

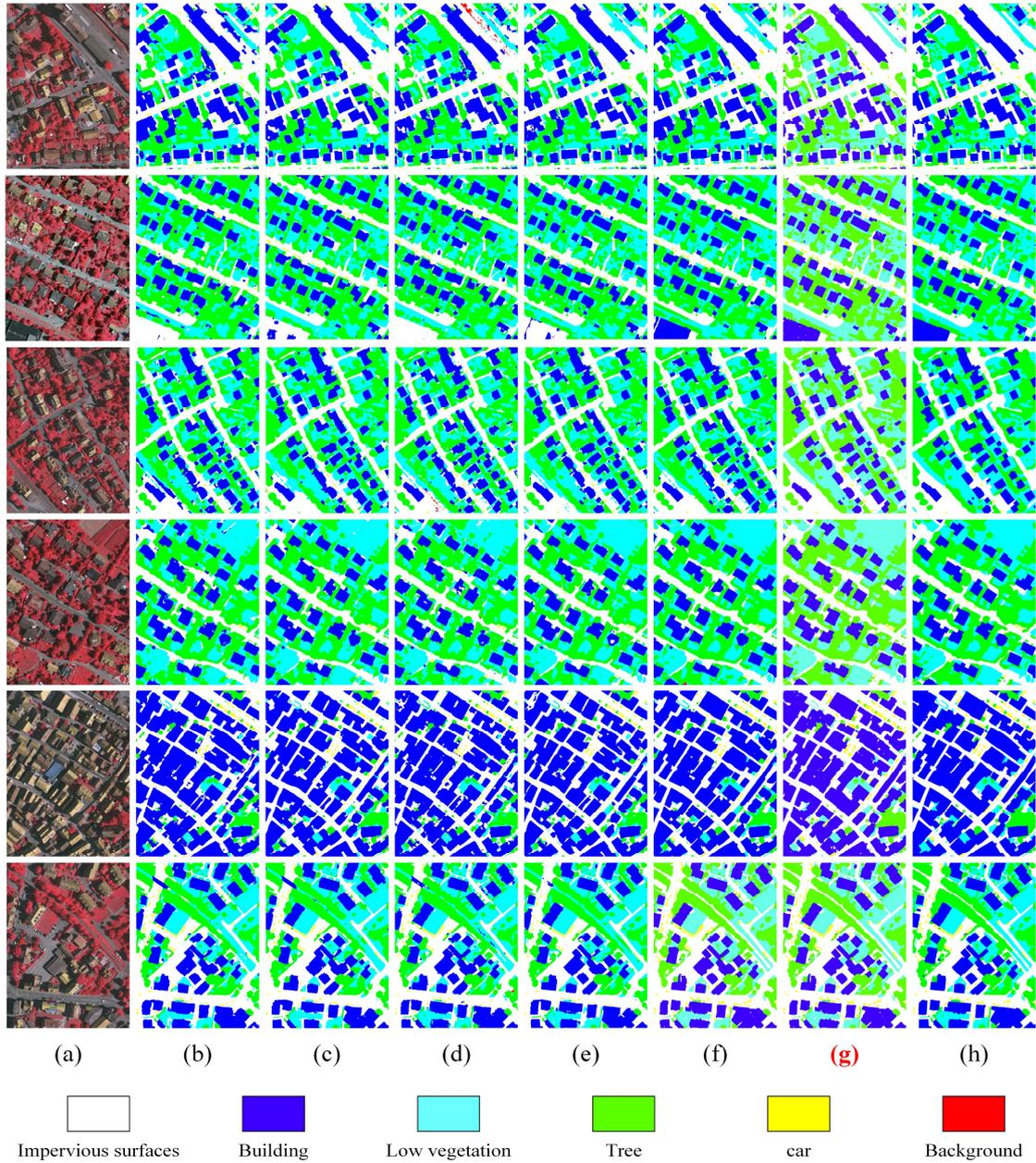

| (a) | (b) | (c) | (d) | (e) | (f) | **(g)** | (h) |

| Impervious surfaces | Building | Low vegetation | Tree | car | Background |

**Fig. 14.** Qualitative comparison of IKD-Net with the benchmark competitors on ISPRS Vaihingen dataset: (a) Imagery, (b) DST_1, (c) DLR_8, (d) UFMG_4, (e) ONE_7, (f) CASIA2, (g) IKD-Net (ours), and (h) Ground truth.

## 5.4 Results on GRSS DFC 2018 dataset

In this section, we review our experiments on the GRSS DFC 2018 dataset to further demonstrate the superiority of our method. The GRSS DFC 2018 dataset was provided by the Image Analysis and Data Fusion Technical Committee for the 2018 IEEE GRSS Data Fusion Contest (DFC). Table 7 lists the best ranked teams in the data fusion classification challenge track (Xu et al., 2019). The best value under a certain metric is bolded.

**Table 7.** Quantitative comparison of IKD-Net and the top ranked teams on GRSS DFC 2018 dataset.

| Team | Mean Acc | OA | Kappa |
|---|---|---|---|
| XudongKang | 0.7126 | 0.7645 | 0.75 |
| Gaussian | 0.7166 | **0.8078** | **0.80** |
| IPIU | 0.7440 | 0.7923 | 0.78 |
| challenger | 0.7599 | 0.7790 | 0.77 |
| AGTDA | 0.7615 | 0.7979 | 0.79 |
| dlrpba | 0.7632 | 0.8074 | **0.80** |
| IKD-Net (ours) | **0.7822** | 0.7828 | 0.77 |

As indicated in Table 7, our IKD-Net ranked highest on Mean Acc and achieved comparable results on OA and Kappa to that of the best performing approaches. It is worth noting that all the best ranked teams adopt post-processing and some of them further employ object detection techniques, which boosted their accuracy by around 15%. However, we still achieved higher scores than the previous winners of the competition and has been ranked first in the real-time leaderboard for the challenge evaluation until this paper's submission. We therefore conclude that our IKD-Net has shown that it is extremely efficient in information utilization and is able to extract deep features from raw multi-modal data and jointly use them according to their inherent characteristics.

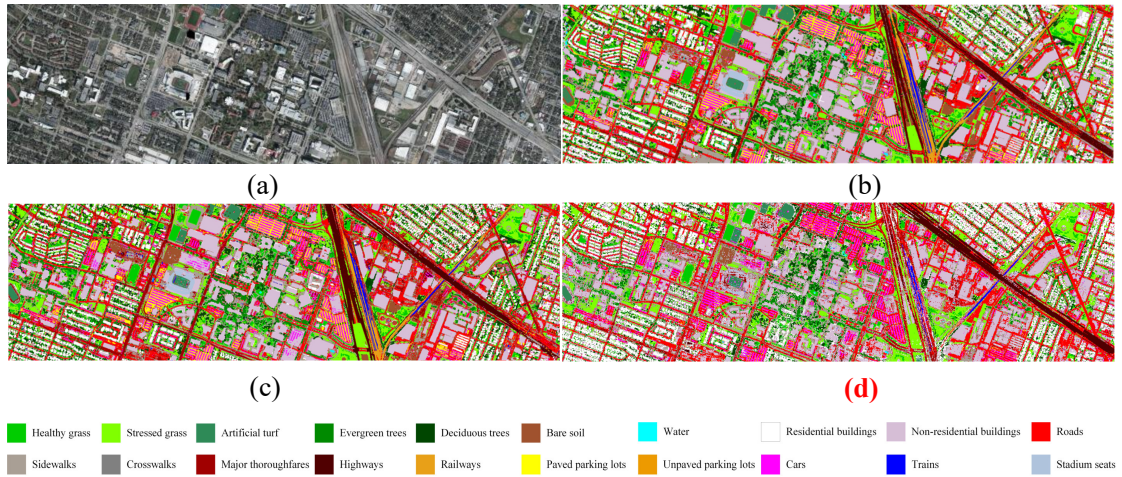

**Fig. 15.** Classification map over the entire scene of GRSS DFC 2018 dataset: (a) Imagery, (b) AGTDA, (c) dlrpba, and (d) IKD-Net (ours).

Fig. 15 shows the imagery and the classification results of the two winning teams and our IKD-Net. Our method excelled in the connectivity of the longest highways (dark-brown) on the right part of the image. There was less confusion between the roads (red) and major thoroughfares (reddish-brown) on our results while team AGTDA and team dlrpba were unable to discriminate these two categories very well, especially on the right side of the images. There were also two obvious minor differences in our approach compared to the top two methods. First, some of the pixels located in the non-residential buildings (lavender) were misclassified as roads (red); and second, some pixels of paved parking lots (yellow) were confused with those of cars (pink).

## 6.   Conclusion

In this paper, we proposed a novel end-to-end heterogeneous dual-stream architecture network called IKD-Net for multi-modal land-cover segmentation. Unlike the current mainstream multi-modal approaches in remote sensing, our dual-stream architecture extracts the features from raw multi-modal heterogeneous data directly rather than their abridged derivatives to retain the intact information of both modalities. Our two GKG and CKG plug-and-play gated modules then utilize the strong modal (LiDAR) to drive the feature map refinement of the weak modality (image) in the global and categorical perspective. The whole network is finally optimized by a sophisticated holistic loss function. In the course of our work, we also established a new dataset called N3C-California to provide a particular benchmark for multi-modal joint segmentation to address the lack of large-scale annotated LiDAR-imagery datasets dedicated to remote sensing tasks. We conducted not only sufficient ablation studies of the above modules and visualized their effects in this paper but conducted additional experiments as well that demonstrated IKD-Net's ability to exceed the benchmarks and the SOTA methods on the N3C-California and ISPRS Vaihingen datasets. Furthermore, IKD-Net has been ranked first in the real-time leaderboard on the GRSS DFC 2018 challenge evaluation until this paper's submission.

In subsequent studies, we aim to enhance the synergy of more modalities and design multi-task networks. Moreover, we also plan to work on the integration of point-clouds and non-ortho-rectified images, such as raw frame aerial images or line-array satellite images to enable more precise mapping of features onto the image-space.

## Acknowledgment

This work was supported by the National Natural Science Foundation of China (Grants 42030102, 42192583, and 42001406), the Fund for Innovative Research Groups of the Hubei Natural Science Foundation (Grant 2020CFA003), the China Postdoctoral Science Foundation (Grant 2020M672416), and the Major special projects of Guizhou [2022]001.